\documentclass[intlimits,twoside,a4paper]{article}

\usepackage[cp1251]{inputenc}

\usepackage[eqsecnum]{cmpj3}

\issue{2024}{27}{2}{23704}
\doinumber{10.5488/CMP.27.23704}

\title[Hexagonal core--shell structure nanowire with high spins of spin-3/2 and spin-5/2]%
{Hexagonal core--shell structure nanowire with high spins of spin-3/2 and spin-5/2}%

\author[M. Karimou,  R. A. Yessoufou, E. Albayrak  ]{M. Karimou\orcid{0000-0002-1422-6479}\refaddr{label1,label2}\thanks{Corresponding author: \email{mounirou.karimou@yahoo.fr}.},
 R. A. Yessoufou\refaddr{label2,
 label3},
 E. Albayrak\refaddr{label4}
 }
\addresses{
\addr{label1} Ecole Nationale Sup\'{e}rieure de G\'{e}nie Energ\'{e}tique et Proc\'{e}d\'{e}s d’Abomey, Benin
\addr{label2} Institute of Mathematic and Physical Sciences (IMSP), Dangbo, Benin
\addr{label4}  Department of Physics, University of Abomey-Calavi, Benin
\addr{label3} Department of Physics, Erciyes
 University, 38039, Kayseri, Turkey
}


\date{Received February 08, 2024, in final form May 23, 2024}

\begin{document}

\maketitle

\begin{abstract}
The magnetic properties and hysteresis loops for the hexagonal Ising nanowire (HIN) with a core--shell structure consisting of mixed spins with the core spin being spin-5/2 and the shell spins being spin-3/2 are studied. The Blume--Capel model is considered by using the mean-field approximation (MFA) based on the Gibbs--Bogoliubov inequality for free energy. The impact of different bilinear interaction parameters ($J_{cc}, J_{ss}, J_{cs}$) between the core, shell, and core and shell spins, respectively, including the crystal ($D_c, D_s$) and external magnetic fields ($h=h_c=h_s$) at the core and shell sites, are taken into consideration.
In order to obtain phase diagrams on various planes, the thermal changes of the net, core, and shell magnetizations are investigated for various values of our system parameters. It is discovered that the model exhibits only second-order phase transitions when $h=0.0$ for $D$ greater or equal to zero, first- and second-order phase transitions for $h\neq0.0$ and compensation temperatures for all $h$.

\printkeywords
%
\end{abstract}

\section{Introduction}
It is well known that as the considered spins get higher, the magnetic behaviors of the materials become richer, displaying various critical phenomena. Furthermore, their mixtures induce extra critical behaviors, such as the compensation temperature at which net magnetization becomes zero before the critical temperatures. Therefore, they have been investigated under various circumstances, following a variety of approaches.

A mixture of high spins with spin-3/2 and spin-5/2 displaying interesting critical features was considered in some works. The phase diagrams and internal energies of the ferrimagnetic Ising system with interlayer coupling were studied by the effective field theory (EFT) with correlations~\cite{Zhang}. In order to study the critical behaviors of the Blume--Capel (BC) Ising ferrimagnetic system, the exact recursion relations were considered on the Bethe lattice with various coordination numbers~\cite{Albayrak}. The exact recursion relations were applied to study the magnetic properties of the BC Ising ferromagnetic system on the two-fold Cayley tree that consists of two sublattices A and B~\cite{Rachidi}. The magnetic properties of a mixed spin-3/2 and spin-2 and a mixed spin-3/2 and spin-5/2 Ising ferromagnetic system with different anisotropies were studied by means of mean-field theory (MFT) and the dependence of the phase diagram on single-ion anisotropy strengths was examined~\cite{Wei}. The dynamic phase transitions and compensation temperatures were examined within the MFA for the Ising system with a crystal field interaction under a time-varying magnetic field on a hexagonal lattice by using Glauber type stochastic dynamics~\cite{Deviren}. The magnetizations of a multilayer thin film described by the transverse Ising model were investigated within the framework of the EFT with correlations, and the influence of the surface exchange coupling and transverse fields on the magnetization and phase transition temperatures was discovered~\cite{Ma}. The ferrimagnetic Ising model with nearest neighbor (NN) interactions was studied in the MFA for both square and simple cubic lattices, and the equilibrium magnetizations and the compensation temperatures were obtained~\cite{Mohamad}. The critical and compensation temperatures of a ferrimagnetic Ising system with mixed spins, $S_i^A= 3/2$ and $\sigma_j^B = 5/2$, were analyzed using Monte Carlo (MC) simulations~\cite{Espriella}. The magnetic properties of a ferrimagnetic Ising model with two crystal fields in a longitudinal magnetic field were studied by MC simulations~\cite{Reyes}. The magnetic properties of the Ising ferrimagnetic system in a graphene layer were explored by means of MC simulations, and the effects of next-nearest neighbors exchange interactions and crystal field anisotropy on the critical and compensation behavior of the system were investigated~\cite{Cardona}. The MC simulation techniques were carried out to study the magnetic behaviors of a mixed Ising system on a square lattice, where spin-3/2 alternates with spin-5/2 in two interpenetrating sublattices A and B~\cite{De La}. The dynamic magnetic properties of the Ising bilayer system consisting of the mixed (3/2, 5/2) Ising spins with a crystal field interaction in an oscillating field on a two-layer square lattice were studied using the dynamic mean-field theory based on the Glauber-type stochastic dynamics~\cite{Ertas}.

This mixed-spin model was also considered for a more sophisticated system, such as the core--shell structured nanowires. Experimental research on core--shell magnetic nanowires made by fusing several magnetic materials was conducted in the recent years~\cite{Nano}. The benefit of these materials is that they convey multifunctional capabilities due to the integrated properties of multiple materials. Additionally, due to this mixing, the core--shell structure can exhibit compensating phenomena.

This type of core--shell structured nanowire has also attracted theoretical interest. The dielectric properties of Ising ferrielectric nanowires with a spin-3/2 core and a spin-5/2 shell structure were systematically studied using the MC simulation in the presence of the external longitudinal electric field. The specific heat, compensation points, susceptibility, and hysteresis behaviors were examined~\cite{Benhouria}. The MC simulations based on the metropolis algorithm were performed to study the critical and compensation temperatures of a core--shell nanowire with mixed spin-3/2 and 5/2, respectively, for an Ising antiferromagnetic system, and the influence of nearest neighbor exchange interactions and crystal field anisotropy on the critical and compensation behaviors of the system was analyzed~\cite{Alzate-Cardona}. The magnetic properties and hysteresis behavior of a ferrimagnetic cubic Ising nanowire with mixed spin-3/2 and spin-5/2 in which the atoms are placed alternately by the MC simulation to investigate the effects of the exchange interactions and crystal field on the magnetic properties and hysteresis behavior of the nanowire~\cite{Aharrouch}. Again, the MC simulation was carried out to investigate the dynamic magnetic behaviors of the ferrimagnetic mixed spin (3/2, 5/2) Ising-type borophene nanoribbons with core--shell structure. The effects of the crystal field, exchange couplings and time-dependent oscillating magnetic field on the dynamic magnetic characteristics were discussed~\cite{Gao}. The phase diagrams and hysteresis loops of a ferrimagnetic mixed spin (3/2, 5/2) hexagonal Ising nanotube with core--shell structure within the framework of MC simulation based of the Metropolis algorithm was applied and the effects of the longitudinal crystal field of the shell sublattice, the exchange couplings were studied on the magnetization behaviors, magnetic susceptibility curves, phase diagrams, and hysteresis loops~\cite{Nmaila, Nmaila1}. A cylindrical Ising nanotube that consists of 3/2 core spins surrounded by 5/2 shell spins was introduced and studied with the MFA and MC simulations in the presence of crystal and external magnetic fields~\cite{Karimou}.

It should be noted that the MFA by using the Gibbs--Bogoliubov inequality was carried out for the nanostructure models including spin-1 and spin-1/2 such as for the magnetic properties of an hexagonal nanosystem with the molecular field-type calculation and MC simulations~\cite{RGB1}, for the thermodynamic states of the hexagonal nanotube system obtained from an eighteen-site cluster within an improved MFA~\cite{RGB2} and for  a seven-site cluster within an improved MFA~\cite{RGB3}.

Note also that the MFA is well recognized to provide just a rough depiction of the true situation while overestimating particle interaction. Efforts to improve the homogeneous the MFA are always increasing. Since the Gibbs--Bogoliubov inequality states that the free energy of a system is always less than that computed by a trial function, the chosen trial function prevents the MFA from being overestimated, resulting in an improvement.

It is clear from the above references that the mixed spin with spin-3/2 and 5/2 nanowire model is only examined using the MC simulations, to the best of our knowledge. Therefore, in this work, we consider another method. Thus, the hexagonal Ising nanowire with a core--shell structure consisting of mixed spins, with the core being spin-5/2 and the shell spins being spin-3/2 is studied for the BC model using the MFA based on the Gibbs--Bogoliubov inequality for free energy.

The rest of the work is organized as follows: section 1 presents the model and the method MFA based on the Gibbs--Bogoliubov inequality for free energy. Section 2 contains all our findings in terms of thermal variations of magnetizations, phase diagrams, and hysteresis loops. The final section includes a brief summary, comparisons, and discussions.

\section{The model and method}

\begin{figure}[htb]
\centerline{\includegraphics[width=0.25\textwidth]{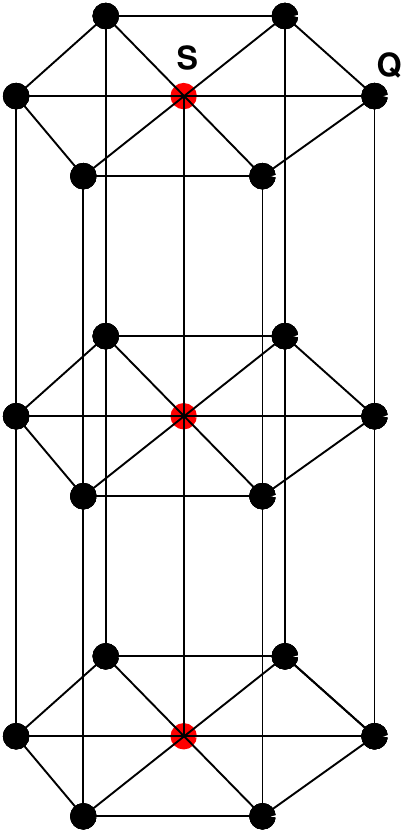}}
\caption{(Colour online) Schematic representation of spin-5/2 chain with hexagonal spin-3/2 shell. The red and black circles represent the magnetic atoms of spin $S$ = 5/2 in the core and spin $Q$ = 3/2 on the surface shell, respectively.} \label{fig-smp1}
\end{figure}

The model under consideration is a mixed spin model with spins 3/2 and 5/2 in a hexagonal nanowire, as schematically depicted in  figure~\ref{fig-smp1}. Each hexagon has one core-spin $S=\frac{5}{2}$ which is enclosed by six shell-spins $Q=\frac{3}{2}$. The hexagonal wire is chosen to have $N=7$ layers and thus consists of a total of $N_{T} = 7 N$ spins.

The Hamiltonian of the model may be written as follows:
\begin{eqnarray}
\label{delta-def}
	H &=&- J_{cs}\sum_{\langle{i,j}\rangle}{S_{i}Q_{j}}
	- J_{c}\sum_{\langle{m,n}\rangle}{S_{m}S_{n}}
	- J_{s}\sum_{\langle{i,m}\rangle}{Q_{i}Q_{m}}
	- D_{c}\sum_{{i}}{(S_{i})^2}\nonumber\\
	&-&  D_{s}\sum_{{m}}{(Q_{m})^{2}}- h\sum_{i}{S_{i}} - h\sum_{m}Q_{m},
\end{eqnarray}
where $S_i$ is a spin-5/2 having the values of $\pm 5/2$, $\pm 3/2$, $\pm 1/2$ and $Q_j$ is spin-3/2 with the values of $\pm3/2$, $\pm1/2$. The first sum runs over all pairs of the NN sites between the $S$--$Q$ spins (core--shell interactions), the second one runs over all core spins of NN hexagons between the $S$--$S$ spins (core--core interactions), and the third sum contains the interactions of the shell spins between the NN hexagons of $Q$--$Q$ spins (shell--shell interactions). Thus, the bilinear exchange interaction parameters $J_{cs}$, $J_{c}$ and $J_{s}$ describe the couplings between the $S$--$Q$, $S$--$S$ and $Q$--$Q$, respectively. $D_{c}$ and $D_{s}$ are the crystal field parameters active at the sites of core and shell spins, respectively. $h$ is the external magnetic field taken to be equal at each lattice sites.

The approximated free energy of the system can be obtained from a variational principle using the Gibbs--Bogoliubov inequality given as
\begin{eqnarray}
\label{delta-def1}
	G(H) \leqslant G(H_0) + \langle H - H_0 \rangle,
\end{eqnarray}
where $G(H)$ is the free energy of the system described by the Hamiltonian given in
equation~(\ref{delta-def}), i.e., the exact free energy. $G(H_0)$ is the average free energy of the trial Hamiltonian $H_0$ which depends on variational parameters and $\langle H - H_{0}\rangle$ denotes the thermal average of the value $H - H_0$ over the ensemble defined by $H_0$.

In the calculations, we have followed the conventional procedure of~\cite{GBI}. The trial Hamiltonian is assumed to be in the form given as
\begin{eqnarray}
	H_{0}=-\sum_{i}\big(\lambda_{c} S_{i}+ D_c S_{i}^{2}\big) -\sum_{m} \big(\lambda_{s} Q_{m}+ D_s Q_{m}^{2}\big),
\end{eqnarray}
where $\lambda_c$ and $\lambda_s$ are the two variational parameters related to the molecular field acting on the core and shells of the hexagonal nanowire.

After some straightforward calculations, the expression for the variational free energy $g$ of equation~(\ref{delta-def1}) is obtained as
\begin{eqnarray}
\label{delta-def2}
g &=& -\frac{N}{\beta}\bigg\{\ln\bigg[2 \exp{\left(\frac{25\beta D_{c}}{4}\right)}\cosh\left(\frac{5}{2}\beta\lambda_{c}\right)+2 \exp{\left(\frac{9\beta D_{c}}{4}\right)}\cosh\left(\frac{3}{2}\beta\lambda_{c}\right)
\nonumber\\
&+& 2 \exp{\left(\frac{\beta D_{c}}{4}\right)}\cosh\left(\frac{1}{2}\beta\lambda_{c}\right)
\bigg]\bigg\}
-\frac{6 N}{\beta}\bigg\{\ln\bigg[
2 \exp{\left(\frac{9\beta D_{s}}{4}\right)}\cosh\left(\frac{3}{2}\beta\lambda_{s}\right)  \nonumber\\
&+&  2\exp{\left(\frac{\beta D_{s}}{4}\right)}\cosh\left(\frac{1}{2}\beta\lambda_{s}\right)\bigg]\bigg\} 
+ (-2J_{c} M_{c}-6J_{cs} M_{s} - h + \lambda_{c}) M_{c} \nonumber\\
&+& (-4J_{s} M_{s}-J_{cs} M_{c}-h +\lambda_{s}) M_{s}.
\end{eqnarray}
The parameters $\lambda_c$ and $\lambda_s$ are found by the minimization procedure, i.e., taking the derivatives of equation~(\ref{delta-def2}) with respect to  $\lambda_{c}$ and $\lambda_{s}$ and equating zero leads to our desired parameters. Thus, one obtains
\begin{eqnarray}
\lambda_{c}&=& 2 J_{c} M_{c} + 6 J_{cs} M_{s} + h,\nonumber\\
\lambda_{s}&=& 4 J_{s} M_{s}+ J_{cs} M_{c} + h.
\end{eqnarray}
As the final equations, the magnetizations of the core and shell spins may be obtained by using $M_{c}=-\frac{1}{N} \frac{\partial g}{\partial \lambda_c}$ and $M_{s}=-\frac{1}{N} \frac{\partial g}{\partial \lambda_s}$, which are calculated respectively as
\begin{eqnarray}
M_{c}=\frac{5\sinh\big(\frac{5}{2}\beta\lambda_{c}\big) + 3\re^{-4\beta D_{c}}\sinh\big(\frac{3}{2}\beta\lambda_{c}\big)+\re^{-6\beta D_{c}}\sinh\big(\frac{1}{2}\beta\lambda_{c}\big) }
{2\cosh\big(\frac{5}{2}\beta\lambda_{c}\big) +
2\re^{-4\beta D_{c}}\cosh\big(\frac{3}{2}\beta\lambda_{c}\big) + 2\re^{-6\beta D_{c}}\cosh\big(\frac{1}{2}\beta\lambda_{c}\big) },
\end{eqnarray}
and
\begin{eqnarray}
M_{s}=\frac{3\sinh\big(\frac{3}{2}\beta\lambda_{s}\big) +\re^{-2\beta D_{s}}\sinh\big(\frac{1}{2}\beta\lambda_{s}\big) }
{2\cosh\big(\frac{3}{2}\beta\lambda_{s}\big) +
2\re^{-2\beta D_{s}}\cosh\big(\frac{1}{2}\beta\lambda_{s}\big) }.
\end{eqnarray}
It should also be mentioned that the average total magnetization per site of an hexagon is given as
\begin{eqnarray}
\label{delta-def3}
|M_{T}|=\frac{M_{c}+6  M_{s}}{7},
\end{eqnarray}
which was investigated in detail by varying our system parameters to calculate the thermal variations and to obtain the phase diagrams in the next section.

\section{The characteristics of magnetizations and the phase diagrams}

In this section, the thermal variations of the core ($M_c$), shell ($M_s$), and total average ($M_T$) magnetizations are examined when the external magnetic field is turned on and off. Their behaviors enable us to obtain the possible phase transition temperatures, i.e., the phase diagrams of the model. Additionally, the magnetic hysteresis loops are also examined.

\subsection{Thermal variations of magnetizations  and hysteresis loops}

\begin{figure}[htb]
\centerline{\includegraphics[width=0.5\textwidth]{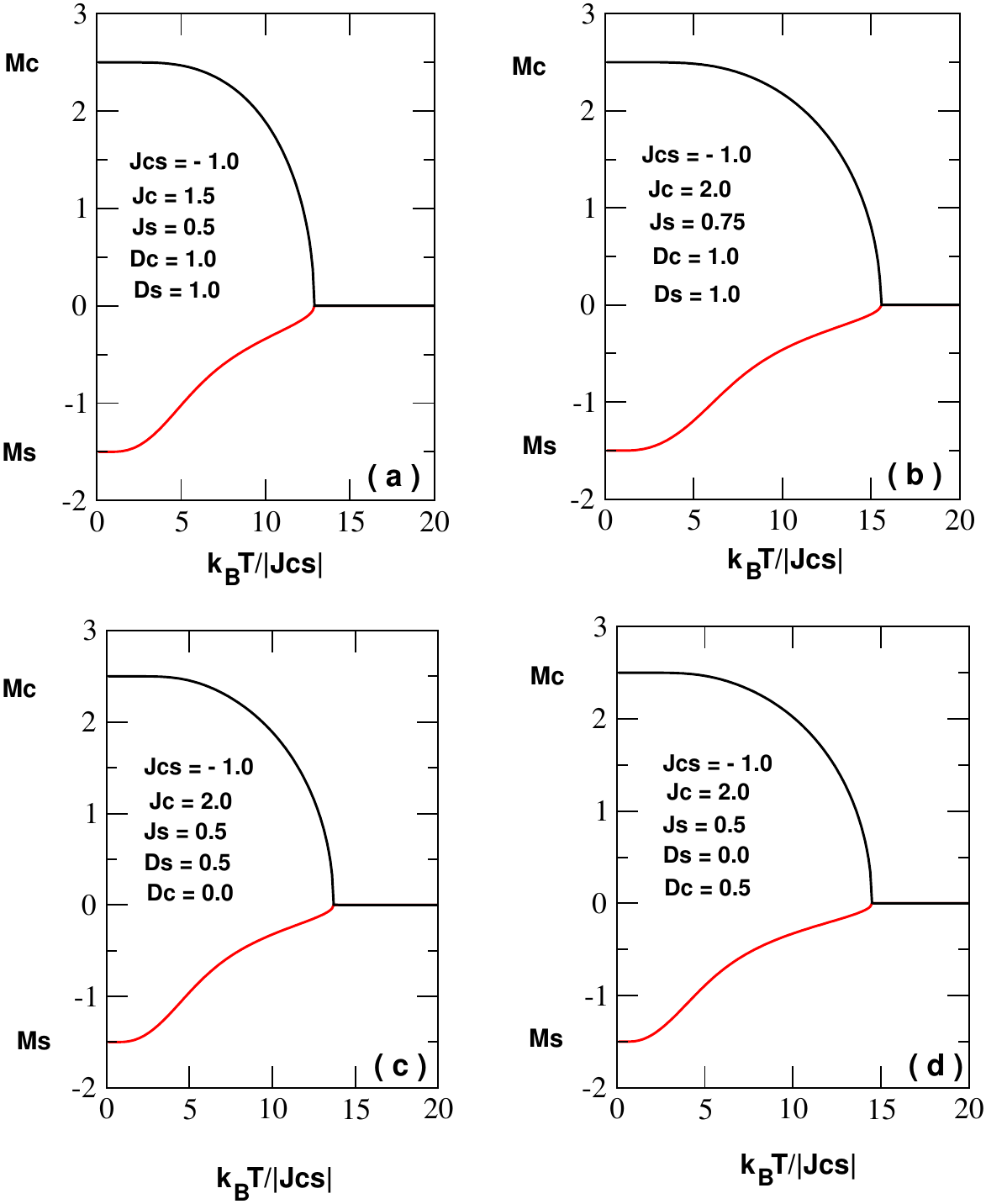}}
\caption{(Colour online) Thermal variations of the core and shell magnetizations $M_{c}$ and $M_{s}$.} \label{fig-smp2}
\end{figure}

As our first illustrations, we have investigated the thermal variations of the core and shell magnetizations, $M_c$ and $M_s$ respectively, with the external magnetic field turned off. It is also assumed that $J_{cs}<0.0$ corresponds to the AFM interactions between the core and shell spins and,  $J_{c}>0.0$ and $J_{s}>0.0$ are for the FM interactions between the shell and between the core spins, respectively. Thus, the core and shell spins are oriented in an antiparallel fashion, as shown in figure~\ref{fig-smp2}(a)--(d). The ground states (GS) of $M_c$ and $M_s$ are 5/2 and $-3/2$ as expected. As the temperature increases, they both decrease to terminate at the common second-order phase transition temperature, i.e., $T_c$. In figure~\ref{fig-smp2}(b), the values of $J_{c}=2.0$ and $J_{s}=0.75$ are larger than the corresponding values in figure~\ref{fig-smp2}(a). Thus, the magnetizations persist at higher temperatures, leading to higher $T_c$. In figure~\ref{fig-smp2}(c) and (d), the values of the crystal field $D_c=0.0$ and $D_s=0.5$ are respectively exchanged, showing that $D_c$ is more dominant since magnetizations endure higher temperatures and thus a higher $T_c$ appears again.

\begin{figure}[htb]
	\centerline{\includegraphics[width=0.45\textwidth]{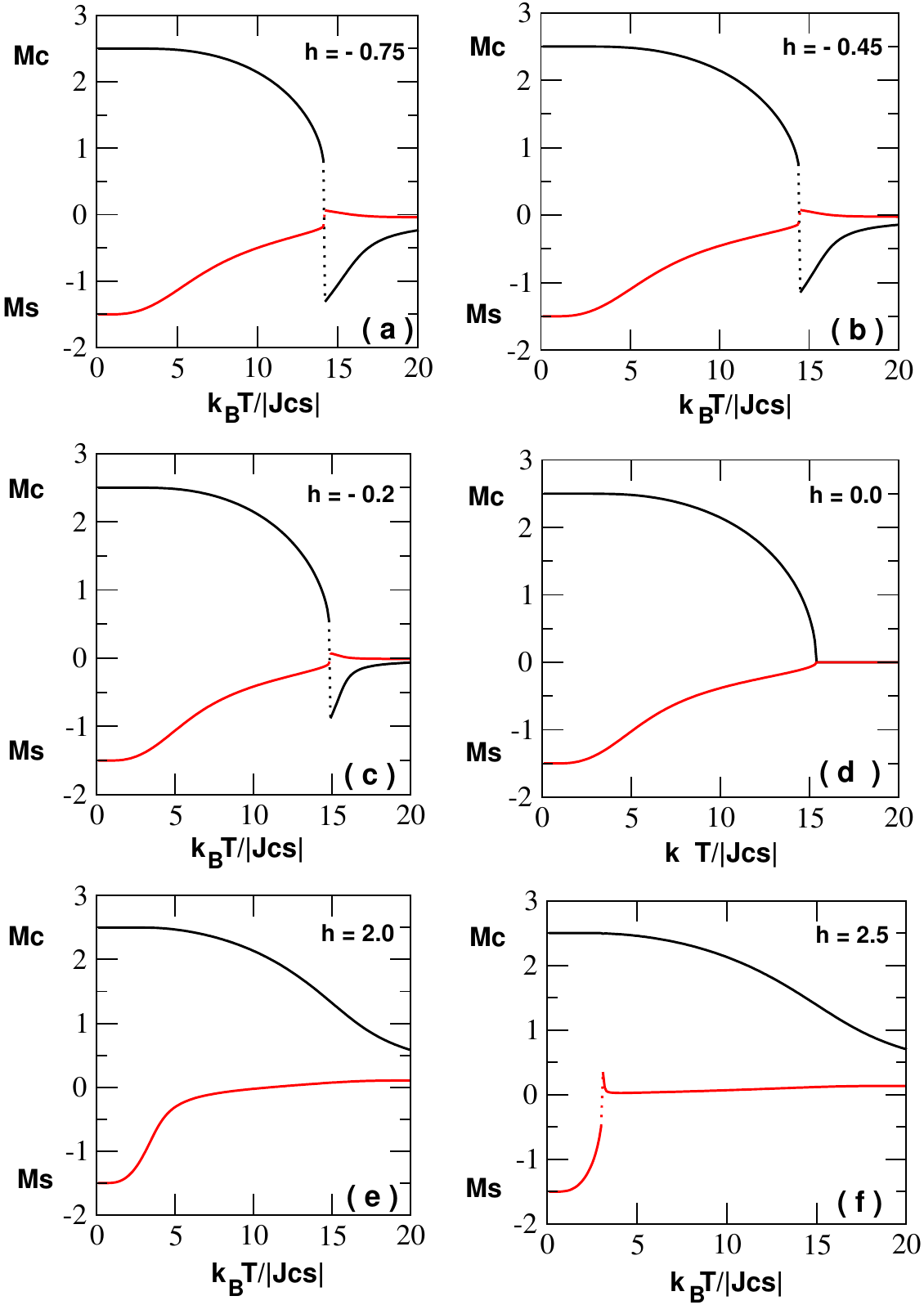}}
	\caption{(Colour online) Thermal variation of the core and shell magnetizations $M_{c}$ and $M_{s}$ with $J_{cs} = - 1.0$, $J_{c}= 2.0$, $J_{s}= 0.5$, $D_{c}=D_{s}= 1.0$ for selected values of $h$.} \label{fig-smp3}
\end{figure}

Figure~\ref{fig-smp3} illustrates the effects of an external magnetic field on the thermal changes of $M_c$ and $M_s$. When $h<0.0$, as shown in figure~\ref{fig-smp3}(a)--(c), both magnetizations start  from their GS values. As temperature increases, they decrease to present jumps at the first-order phase transition temperatures, $T_t$. While $M_s$ jumps from negative to positive values, $M_c$ does the opposite. The jumps get less pronounced as $h$ becomes less negative. At $h=0.0$, as shown in figure~\ref{fig-smp3}(d), both magnetizations vanish at the $T_c$. Figure~\ref{fig-smp3}(e) is obtained for $h=2.0$ and does not present any phase transitions. The final characteristic behavior is displayed in figure~\ref{fig-smp3}(f) where $M_s$ presents a $T_t$ but $M_c$ does not show any phase transitions. Note also that all the figures with $h\neq0.0$ show that magnetizations approach the zero asymptoticaly at higher temperatures.

\begin{figure}[h!]
\centerline{\includegraphics[width=0.5\textwidth]{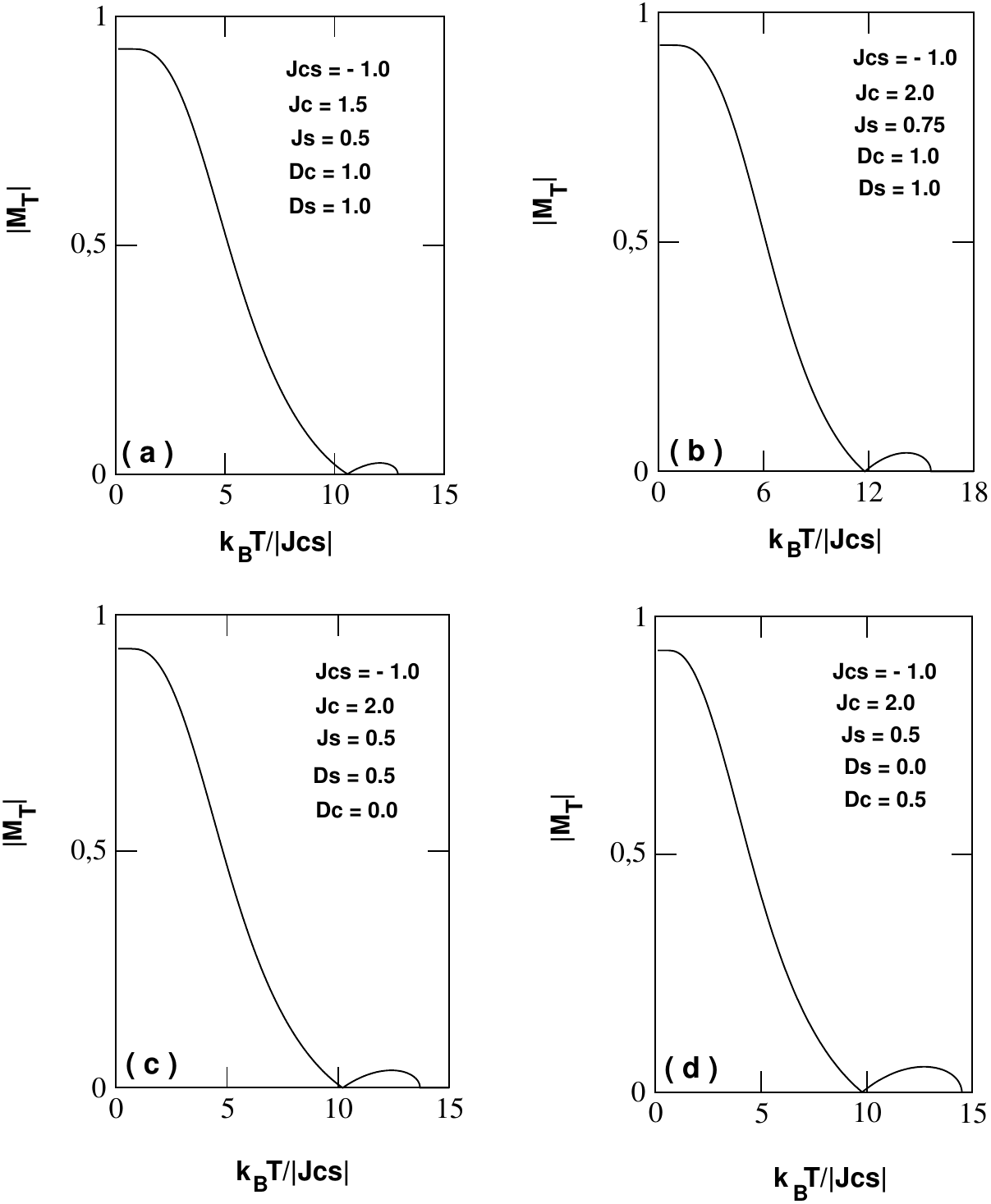}}
\caption{The average total magnetization $|M_{T}|$ versus the temperature.} \label{fig-smp4}
\end{figure}

Figure~\ref{fig-smp4} is obtained in correspondence with figure~\ref{fig-smp2} for $M_T$. According to
equation~(\ref{delta-def3}), $M_T$ contains six $M_s$ and one $M_c$ for each hexagon which must be kept in mind for understanding the figures.  The GS of  $|M_T|=|(2.5-6\times1.5)/7|\simeq0.9285$ since $J_{cs}<0.0$, thus, $ |M_T| $ has this value at zero temperature. The compensation temperature, $T_{\text{comp}}$, is the temperature that appears before $T_c$ and the magnetizations of six shell spins compensate the magnetization of the core spin. It is clear from figure~\ref{fig-smp4}(a) and (b) that as $J_s$ increases, the $T_c$ and $T_{\text{comp}}$ appear at higher $T$, since when $J_s>0.0$, the FM phase is supported. The same can also be expected for $J_c>0.0$. In the last two figures, the values of $D_s$ and $D_c$ are exchanged. It is clear that when $D_s$ is larger than $D_c$, the critical temperatures are seen at higher $T$. This may be caused by the higher number of shell spins than core spins and by the crystal field being positive.

\begin{figure}[h]
	\centerline{\includegraphics[width=0.55\textwidth]{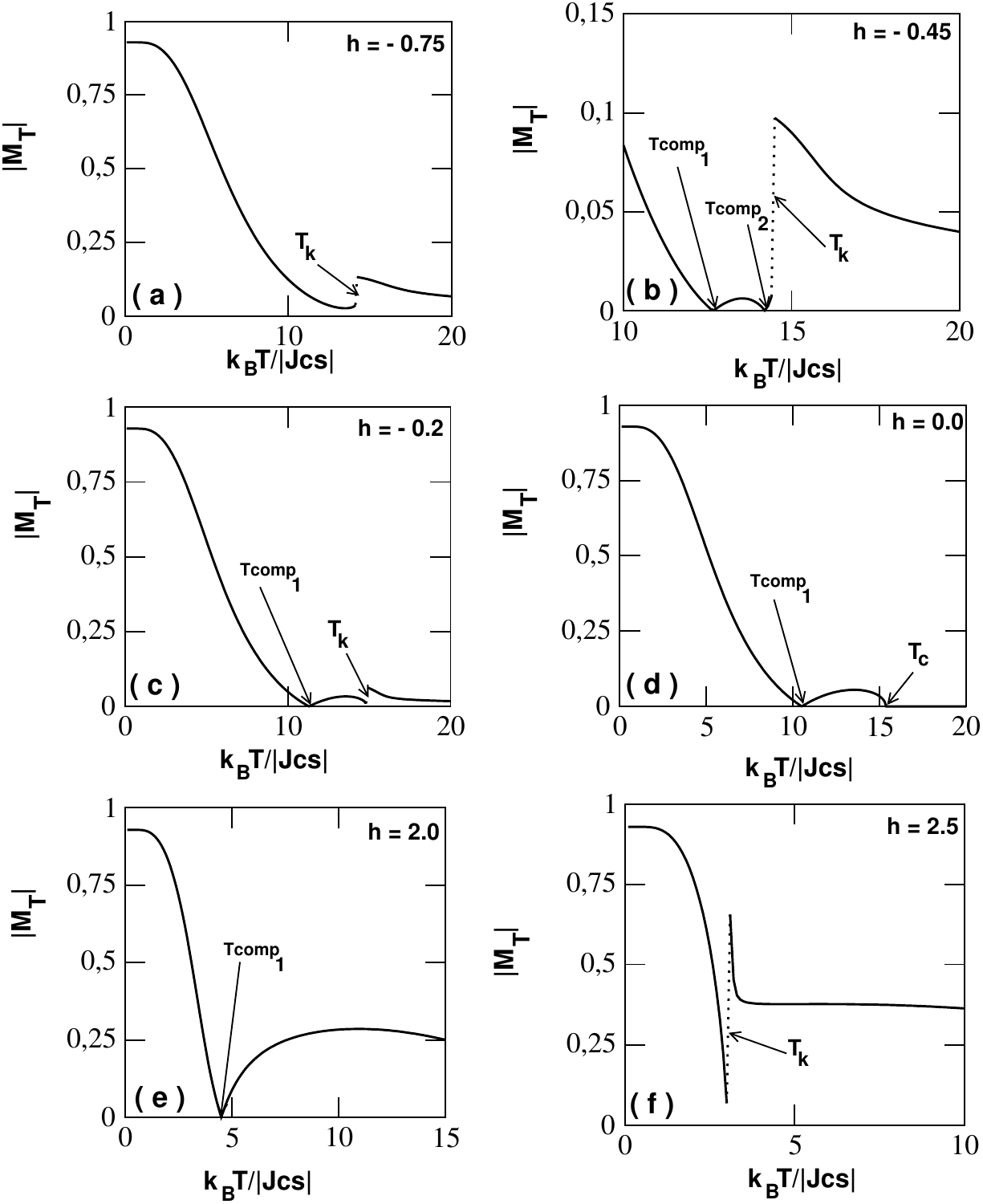}}
	\caption{The average total magnetization $|M_{T}|$ versus the temperature with $J_{cs} = - 1.0$, $J_{c}= 2.0$, $J_{s}= 0.5$, $D_{c}= 1.0$, $D_{s}= 1.0$ for selected values of $h$.} \label{fig-smp5}
\end{figure}

Similarly, figure~\ref{fig-smp5}(a)--(f) is calculated with respect to figure~\ref{fig-smp3}(a)--(f) for the given values of $h=-0.75$, $-0.45$, $-0.2$, $0.0$, $2.0$ and $2.5$. They all start again with the same GS value. The first figure shows one $T_t$, the second one two $T_{\text{comp}}$ and a $T_t$, the third one a $T_{\text{comp}}$ and a $T_t$, the fourth one a $T_{\text{comp}}$ and a $T_c$ (for $h=0.0$), the fifth one only a $T_{\text{comp}}$ and the last one only a $T_{t}$. It is clear that the existence of a $T_{\text{comp}}$ can be accompanied with the phase transitions of any kind or can exist independently. The places of $T_c$ and $T_t$ and the expected positions of $T_{\text{comp}}$ are in agreement with figure~\ref{fig-smp3}, as expected.

\begin{figure}[h]
\centerline{\includegraphics[width=0.55\textwidth]{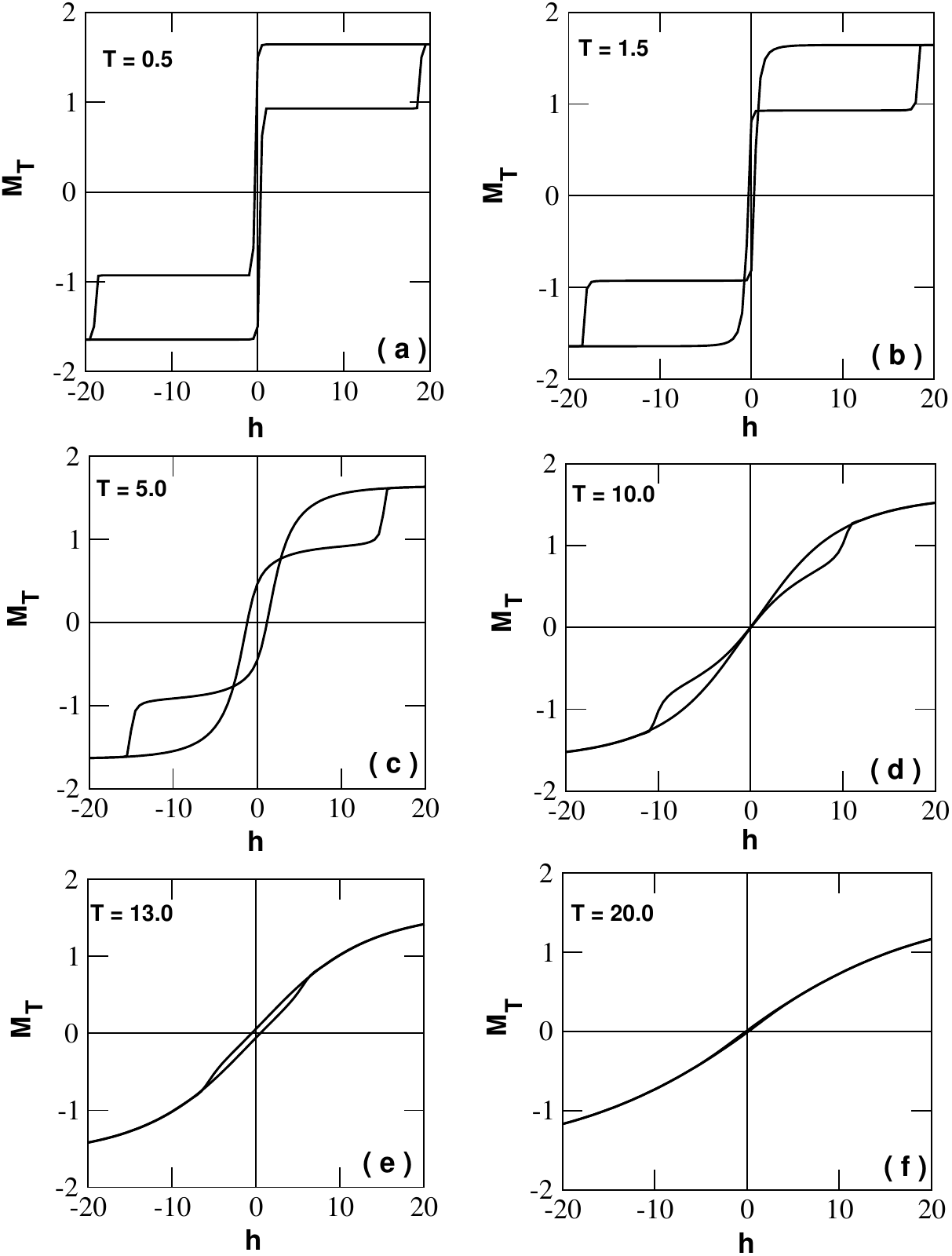}}
\caption{The hysteresis loops with $J_{cs} = - 1.0$, $J_{c}= 2.0$, $J_{s}= 0.5$, $D_{c}= 1.0$, $D_{s}= 1.0$ for selected values of $T$.} \label{fig-smp6}
\end{figure}

In order to investigate the effects of the system parameters on the magnetic hysteresis loops, the total average magnetization curves of the system are plotted as a function of the external magnetic field $h$ for various temperatures, as shown in figure~\ref{fig-smp6}. The first two figures show that $M_T$ for $T=0.5$ and 1.5
have two straight portions in the negative and positive $h$ regions with a sharp rise about $h=0.0$. For $T=0.5$, there is only one cycle since the magnetization lines during the rise do not cross each other. When $T=1.5$, the lines cross each other during the rise which leads to three loops. For $T=5.0$, the three loops become more pronounced, but the straight portions of $M_T$ disappear and appear gradually rising as $h$ rises. As $T$ increases to 10.0, the middle loop disappears, and only two loops remain for $\pm h$ regions. As $T$ increases further to 13.0, these two loops disappear but now concentrate as one loop around $h=0.0$. Finally, all the loops disappear when $T=20.0$.

\subsection{The phase diagrams with $h=0.0$ and $h\neq0.0$}
\begin{figure}[h]
	\centerline{\includegraphics[width=0.55\textwidth]{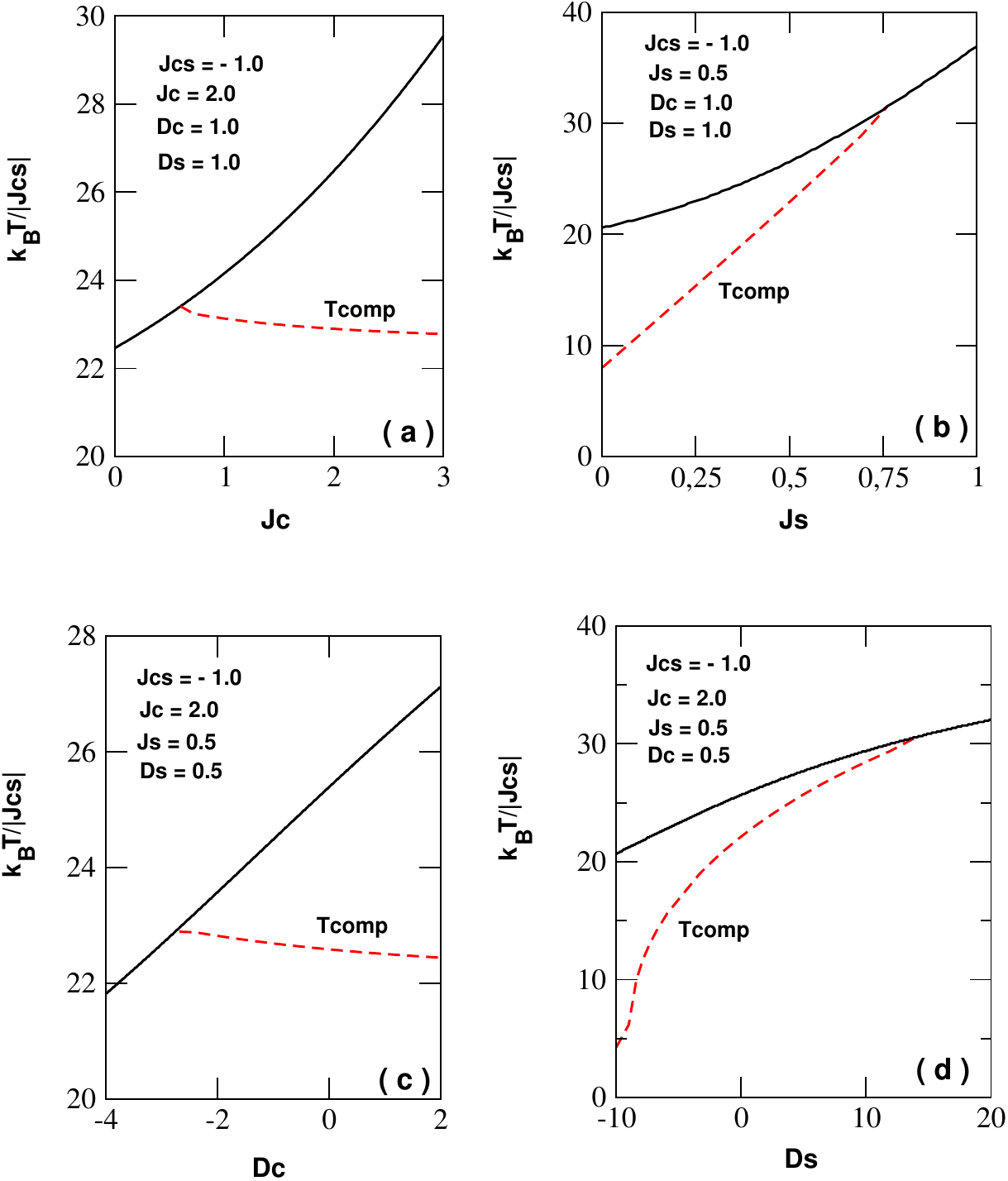}}
	\caption{(Colour online) The phase diagrams on the (a) $(J_{c}, k_{B}T/|J_{cs}|)$, (b) $(J_{s}, k_{B}T/|J_{cs}|)$, (c)~$(D_{c}, k_{B}T/|J_{cs}|)$ and (d) $(D_{s}, k_{B}T/|J_{cs}|)$ planes.} \label{fig-smp7}
\end{figure}

Having investigated the thermal changes of magnetizations under various circumstances, we are now ready to obtain the phase diagrams when $h=0.0$ and $h\neq0.0$ cases:
Figure~\ref{fig-smp7}(a) is obtained for $J_{cs}=-1.0$, $J_s=0.5$ and $D_c=D_s=1.0$ on the ($J_c, k_B T/|J_{cs}|$) plane shows that the $T_c$-line starts at high temperatures for $J_c=0.0$ and increases as $J_c$ increases. This is obvious since increasing $J_c$ values try to align the spins of the cores in favor of the FM phase. The corresponding $T_{\text{comp}}$-line starts on the $T_c$-line and changes very slowly as $J_c$ increases. A similar phase diagram is also obtained on the ($D_c, k_B T/|J_{cs}|$) plane for $J_{cs}=-1.0$, $J_c=2.0$, $J_s=0.5$ and $D_s=0.5$ as given in figure~\ref{fig-smp3}(c). Now, the $T_c$-line appears to be straight, and the $T_{\text{comp}}$ starts from the negative $D_c$ values. Figure~\ref{fig-smp7}(b) obtained for  $J_{cs}=-1.0$, $J_c=2.0$ and $D_c=D_s=1.0$ on the ($J_s, k_B T/|J_{c}|$) plane and figure~\ref{fig-smp7}(d) calculated for $J_{cs}=-1.0$, $J_c=2.0$, $J_s=0.5$ and $D_c=0.5$ on the ($D_s, k_B T/|J_{cs}|$) plane show different characteristic behaviors. The $T_{\text{comp}}$-lines start immediately with $J_s=0.0$ in figure~\ref{fig-smp7}(b) and with $D_s=0.0$ in figure~\ref{fig-smp7}(d) and rise as $J_s$ and $D_s$ rise, respectively. Then, they combine with the $T_c$-lines. Note also that the $T_c$-line of figure~\ref{fig-smp7}(b) increases at a constant rate as $J_s$ increases further, but for figure~\ref{fig-smp7}(d), the increase of the $T_c$-line slows as $D_s$ increases further. The increase of the critical lines with increasing $J_s$ and $D_s$ in comparison with their corresponding core spin values is clear since each shell contains six spins.

The final illustration of this work, obtained on the ($h, k_B T/|J_{cs}|$) plane for $J_{cs}=-1.0$, $J_c=2.0$, $J_s=0.5$ and $D_c=D_s=1.0$, is quite interesting. It contains a $T_c$-line starting at a high $T$ for $h<0.0$, then decreasing as $h$ increases, then after making a dip, it starts rising with increasing $h$ and finally becomes constant. In addition, a $T_t$-line starts from the lower temperature for $h<0.0$ again, increases with increasing $h$, and then terminates at an end point. The $T_{\text{comp}}$-line is also observed, which presents a reentrant behavior in the $h<0.0$ region, then decreases in temperature with increasing $h$ which then terminates combining with the second portion of a $T_t$-line.  Now this $T_t$-line decreases as $T$ increases. It is clear from the inset that these three lines at higher $T$ do not combine.

\begin{figure}[h!]
\centerline{\includegraphics[width=0.45\textwidth]{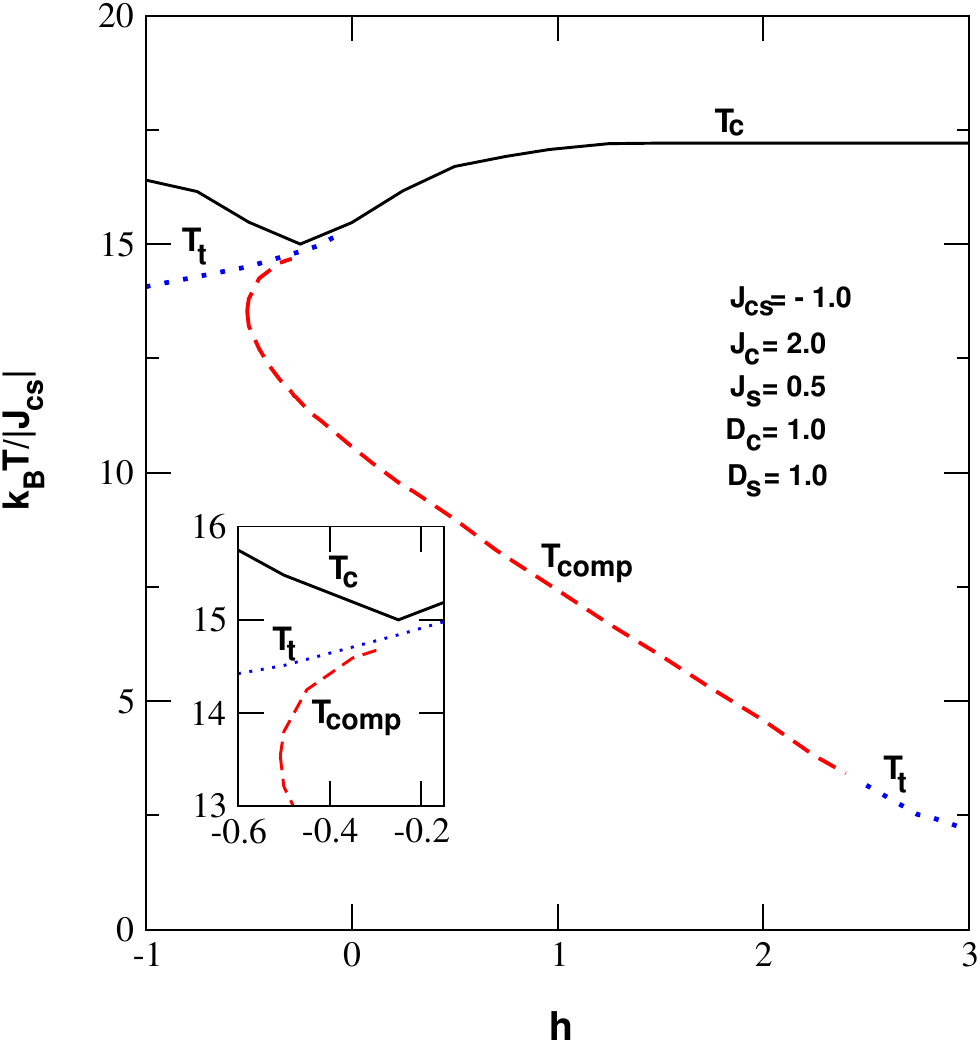}}
\caption{(Colour online) The phase diagram on the $(h, k_{B}T/|J_{cs}|)$ plane.} \label{fig-smp8}
\end{figure}

\section{The summary, conclusions and comparisons}
In this work, the magnetic properties and hysteresis loops of a HIN with a mixed spin core--shell structure are examined. To consider the Blume--Capel model, the MFA based on the Gibbs--Bogoliubov inequality for free energy is utilized. Several bilinear interaction parameters ($J_cc, J_ss, J_cs$) between the core, shell, and core and shell spins, as well as the crystal ($D_c, D_s$) and external magnetic fields ($h=h_c=h_s$) at the core and shell locations, are considered. Thermal variations in the net, core, and shell magnetizations are investigated for various values of our system parameters in order to generate phase diagrams on the ($J_c, k_B T/|J_{cs}|$), ($D_c, k_B T/|J_{cs}|$) and ($h, k_B T/|J_{cs}|$) planes. It is found that the model only exhibits the second-order phase transitions when $h=0.0$ for $D$ greater or equal to zero, first- and second-order phase transitions when $h\neq0.0$, and compensation temperatures for all $h$. The $T_{\text{comp}}$-line for $h \neq 0.0$ presents a reentrant behavior at higher temperatures due to the existence of two compensation temperatures for the given parameters. The calculated hysteresis behavior only displays three loops at maximum, as in~\cite{Nmaila,Benhouria,Karimou}.
Even if there are quantitative differences between this work and the existing ones, we can mention a few works that present very strong qualitative similarities.
Our first phase diagram given in figure~\ref{fig-smp7}(a) is qualitatively similar to the figure~\ref{fig-smp5}(a) of~\cite{Alzate-Cardona}, figure~\ref{fig-smp7} of~\cite{Aharrouch} and figure~\ref{fig-smp3}(a) and~\ref{fig-smp4}(a) of~\cite{Karimou}. Our figure~\ref{fig-smp7}(b) is also qualitatively similar to figure~\ref{fig-smp5}(b) of~\cite{Alzate-Cardona},  figure~\ref{fig-smp7} of~\cite{Nmaila}, figure~\ref{fig-smp8} of~\cite{Nmaila1} and figure~\ref{fig-smp3}(b) and~\ref{fig-smp4}(b) of~\cite{Karimou}. Figure~\ref{fig-smp3}(c)--(d) and figure~\ref{fig-smp4} (c)-(d) of~\cite{Karimou} are also identical with our figure~\ref{fig-smp7}(c)--(d). The phase diagram on the ($h, k_B T/|J_{cs}|$) plane was also presented in figure~\ref{fig-smp2} of~\cite{Alzate-Cardona} for $h>0$ which is similar to our figure~\ref{fig-smp8}. Our $T_c$-line is similar, but the $T_{\text{comp}}$-line declines as $h$ increases, contrary to~\cite{Alzate-Cardona} where it becomes constant.

\ukrainianpart

\title{Нанодріт з гексагональною структурою ``ядро--оболонка'' зі спінами спін-3/2 та спін-5/2}

\author{М. Каріму\refaddr{label1,label2},
	Р. А. Єсуфу\refaddr{label2,label3},
	Е. Албайрак\refaddr{label4}}
\addresses{
	\addr{label1} Національна школа енергетики та інженерії Абомея
	\addr{label2} Інститут математичних та фізичних наук, Дангбо, Бенін
	\addr{label4} Фізичний факультет, Університет Абомея-Калаві, Бенін
	\addr{label3} Фізичний факультет, Університет Ерджіес, 38039, Кайсері, Туреччина
}

\makeukrtitle

\begin{abstract}
	\tolerance=3000%
	Досліджено магнітні властивості та петлі гістерезису для гексагонального нанодроту Ізінга зі структурою ``ядро--оболонка'', що складається зі змішаних спінів, коли спін ядра має дорівнює 5/2, а спін оболонки --- 3/2. Модель Блюма-Капеля розглядається у наближенні середнього поля на основі нерівності Гіббса-Боголюбова для вільної енергії. Враховано вплив різних параметрів нелінійної взаємодії ($J_{cc}, J_{ss}, J_{cs}$) між ядром, оболонкою та спінами ядра і оболонки, відповідно, включаючи кристал ($D_c, D_s$) та зовнішні магнітні поля ($h=h_c=h_s$) у вузлах ядра і оболонки. Щоб отримати фазові діаграми на різних площинах, досліджуються теплові зміни намагніченості сітки, ядра та оболонки при різних значеннях параметрів системи. Виявлено, що модель демонструє лише фазові переходи другого роду, коли $h=0.0$ для $D$ є невід'ємним, а також знайдено фазові переходи першого та другого роду для $h\neq0.0$ та компенсаційні температури для всіх $h$.
	\keywords теорія середнього поля, модель Блюма--Капеля, петлі гістерезису, температура компенсації, намагніченість, система зі змішаними спінами
	
\end{abstract}


\begin{thebibliography}{99}
\bibitem{Zhang}
Zhang Q., Wei G., Gu Y., 	Phys. Status Solidi B, 2005, {\bf{242}}, 924, \doi{10.1002/pssb.200402104}.
\bibitem{Albayrak}
 Albayrak E.,  Yigit A., Phys. Lett. A, 2006, {\bf{353}}, 121,  \doi{10.1016/j.physleta.2005.12.077}.
\bibitem{Rachidi}
Yessoufou R. A., Amoussa S. H., Hontinfinde F., Cent. Eur. J. Phys.,  2009, {\bf{7}}, 555, \doi{10.2478/s11534-009-0016-x}.
\bibitem{Wei}
 Wei G., Hai-Ling M., Commun. Theor. Phys., 2009, {\bf{51}}, 756,  \doi{10.1088/0253-6102/51/4/32}.
\bibitem{Deviren}
Deviren B.,  Keskin M., J. Stat. Phys., 2010, {\bf{140}}, 934,  \doi{10.1007/s10955-010-0025-6}.
\bibitem{Ma}
Ma B., Jiang W., IEEE Trans. Magn., 2011, {\bf{47}}, 3118,  \doi{10.1109/TMAG.2011.2149506}.
\bibitem{Mohamad}
Mohamad H. K., J. Magn. Magn. Mater., 2011, {\bf{323}}, 61,  \doi{10.1016/j.jmmm.2010.08.030}.
\bibitem{Espriella}
De La Espriella Velez N., Ortega Lopez C.,  Torres Hoyos F.,  Rev. Mex. Fis., 2013, {\bf{59}}, 95.
\bibitem{Reyes}
Reyes J. A., De La Espriella N.,  Buend\`{\i}a G. M., Phys. Status Solidi B, 2015, {\bf{252}}, 2268,\\ \doi{10.1002/pssb.201552110}.
\bibitem{Cardona}
Alzate-Cardona J. D., Sabogal-Su\' arez D.,  Restrepo-Parra E., J. Magn. Magn. Mater., 2017, {\bf{429}}, 34,\\ \doi{10.1016/j.jmmm.2017.01.004}.
\bibitem{De La}
De La Espriella N., Madera J. C., Buend\`{\i}a G. M., J. Magn. Magn. Mater., 2017, {\bf{442}}, 350,\\ \doi{10.1016/j.jmmm.2017.07.015}.
\bibitem{Ertas}
Keskin M., Erta\c{s} M., Physica A, 2018, {\bf{496}}, 79,  \doi{10.1016/j.physa.2017.12.034}.
\bibitem{Nano}
Ivanov Y. P., Alfadhel A., Alnassar M., Perez J. E., Vazquez M., Chuvilin A.,  Kosel J.,  Sci. Rep., 2016, {\bf{6}}, 24189,\\ \doi{10.1038/srep24189}.
\bibitem{Benhouria}
Benhouria Y., Essaoudi I.,  Ainane A., Ahuja R., Dujardin F., Ferroelectrics, 2017, {\bf{507}},   58,\\  \doi{10.1080/00150193.2017.1283170}.
\bibitem{Alzate-Cardona} Alzate-Cardona J. D., Barrero-Moreno M. C., Restrepo-Parra E., J. Phys.: Condens. Matter, 2017, {\bf{29}},  445801,\\ \doi{10.1088/1361-648X/aa8a06}.
\bibitem{Aharrouch}
Aharrouch R.,  El Kihel K.,  Madani M., Hachem N., Lafhal A., El Bouziani M., Multidiscip. Model. Mater. Struct., 2020, {\bf{16}}, 1261, \doi{10.1108/MMMS-11-2019-0194}.
\bibitem{Gao}
 Gao Z. Y., Wang W., Sun L., Yang L. M., Ma B. Y., Li P. S.,  J. Magn. Magn. Mater., 2022, {\bf{548}}, 168967,\\  \doi{10.1016/j.jmmm.2021.168967}.
\bibitem{Nmaila}
Nmaila B., Kadiri A., Arbaoui A., Drissi L. B., Ahl Laamara R.,  Htoutou K., Chin. J. Phys., 2022, {\bf{79}}, 362–373,\\  \doi{10.1016/j.cjph.2022.07.019}.
\bibitem{Nmaila1}
Nmaila B., Htoutou K., Ahllaamara R.,  Drissi L. B., Indian J. Phys.,  2023, {\bf{97}}, 429,  \doi{10.1007/s12648-022-02393-1}.
\bibitem{Karimou} Karimou M.,  Oke T. D.,  Hontinfinde S. I. V., Kple J.,  Hontinfinde F.,  Physica B, 2023, {\bf{666}}, 415107,\\ \doi{10.1016/j.physb.2023.415107}.
\bibitem{RGB1} Mendes R. G. B., Santos J. P., S\'{a} Barreto F. C., Braz. J. Phys., 2021, {\bf{51}}, 1929, \doi{10.1007/s13538-021-00982-9}.
\bibitem{RGB2} Mendes R. G. B., S\'{a} Barreto F. C., Santos J. P., Physica A, 2018, {\bf{505}}, 1186, \doi{10.1016/j.physa.2018.03.094}.
\bibitem{RGB3} Mendes R. G. B., S\'{a} Barreto F. C., Santos J. P., Braz. J. Phys., 2018, {\bf{48}}, 137, \doi{10.1007/s13538-018-0560-1}.
\bibitem{GBI} Bogoliubov N. N., J. Phys. (USSR),  1947, {\bf{11}}, 23.
\end{thebibliography}
\lastpage
\end{document}